%% file: ICC2026-1571219380.tex
\documentclass[conference]{IEEEtran}
\IEEEoverridecommandlockouts
\input{preamble}

\begin{document}
\title{System Modeling of Microfluidic Molecular Communication: A Markov Approach}

\author{
    \IEEEauthorblockN{
        Ruifeng Zheng\IEEEauthorrefmark{1},
        Pengjie Zhou\IEEEauthorrefmark{1},
        Pit Hofmann\IEEEauthorrefmark{1},
        Fatima Rani\IEEEauthorrefmark{1},
        Juan A. Cabrera\IEEEauthorrefmark{1}, and
        Frank H.\,P. Fitzek\IEEEauthorrefmark{1}\IEEEauthorrefmark{2}
    }
    \IEEEauthorblockA{
        \IEEEauthorrefmark{1}Deutsche Telekom Chair of Communication Networks, Dresden University of Technology, Germany\\
        \IEEEauthorrefmark{2}Centre for Tactile Internet with Human-in-the-Loop (CeTI), Dresden, Germany\\
        \texttt{\{firstname.lastname\}@tu-dresden.de}
    }
    \thanks{This work was funded by the German Research Foundation (DFG, Deutsche Forschungsgemeinschaft) as part of Germany’s Excellence Strategy – EXC 2050/1 – Project ID 390696704 – Cluster of Excellence “Centre for Tactile Internet with Human-in-the-Loop” (CeTI) of Dresden University of Technology.
    The authors also acknowledge the financial support by the Federal Ministry of Research, Technology and Space (BMFTR) of Germany in the program “Souverän. Digital. Vernetzt.” Joint project 6G-life, project identification number 16KISK001K.
    Furthermore, this work was partly supported by the projects IoBNT (grant number 16KIS1994) and MoMiKoSy (Software Campus, grant number 16S23070), funded by the Federal Ministry of Research, Technology and Space (BMFTR).}
    }

\maketitle


\begin{abstract}
This paper presents a Markov-based system model for microfluidic molecular communication (MC) channels. By discretizing the advection–diffusion dynamics, the proposed model establishes a physically consistent state–space formulation. The transition matrix explicitly captures diffusion, advective flow, reversible binding, and flow-out effects. The resulting discrete-time formulation enables analytical characterization of both transient and equilibrium responses through a linear system representation. Numerical results verify that the proposed framework accurately reproduces channel behaviors across a wide range of flow conditions, providing a tractable basis for the design and analysis of MC systems in microfluidic environments.
\end{abstract}


\section{Introduction} \label{sec:intro}
\IEEEPARstart{M}{olecular} communication (MC) is an emerging bio-inspired paradigm in which chemical signals are used to convey information among nanoscale or microscale devices~\cite{farsad2016comprehensive,huang2019spatial,qiu2023review}. Unlike electromagnetic or acoustic signaling, MC leverages physical processes such as diffusion, advection, and chemical reactions to transmit and process information in aqueous or microfluidic environments~\cite{zheng2025anis}. Such mechanisms are inherently compatible with biological systems, enabling applications ranging from targeted drug delivery and in-body sensing to lab-on-a-chip and DNA-based computation platforms~\cite{zheng2025DNA-Based,xiang2023tutorial}.

Among the broad variety of MC environments, \emph{microfluidic MC} has received growing attention due to its controllable flow conditions, reproducible geometries, and compatibility with biochemical sensing interfaces~\cite{walter2023real,hamidovic2024microfluidic}. In microfluidic channels, \acp{IM} are transported by a combination of advection and diffusion, while the receiver surface often involves specific molecular recognition mechanisms such as ligand–receptor or DNA hybridization reactions. Accurately modeling these coupled advection-diffusion dynamics is crucial for characterizing channel behavior and optimizing system performance. 

Traditional analytical models for IM transport are typically based on partial differential equations (PDEs) describing diffusion and advection with reactive boundary conditions~\cite{wicke2019magnetic,jamali2019channel}. While such formulations capture the physical phenomena precisely, their continuous nature and spatial complexity often lead to intractable or numerically expensive solutions, particularly when reversible surface reactions or flow-out effects are included. This motivates the development of \emph{discrete, and computationally tractable} models that approximate the underlying dynamics while preserving key physical properties, such as probability conservation and equilibrium consistency.

To address this challenge, we propose a \emph{Markov-based system model} for microfluidic MC channels. By discretizing both the spatial domain and surface-reaction kinetics, the proposed framework reformulates the advection-diffusion process into a discrete-time, discrete-state Markov chain. The transition matrix explicitly encodes diffusion, advection, reversible binding, and flow-out mechanisms, providing a compact and physically consistent representation of \ac{IM} propagation. Building on this Markov framework, we derive a linear time-domain system model that links the transmitter input to the receiver observation through a channel impulse response (CIR), enabling analytical characterization of both transient and equilibrium behaviors. 

The proposed model is applicable to a wide range of microfluidic scenarios where the transport and surface reactions of IMs are coupled. Typical examples include DNA microarray hybridization, surface-based affinity assays, and microfluidic biosensors, in which diffusion, flow, and binding jointly determine the sensing response~\cite{hassibi2005biological,squires2008making}. Beyond biosensing, the model can also support system-level analysis of MC channels and lab-on-chip drug delivery systems~\cite{wu2018lab}, providing an analytically tractable framework for performance evaluation and control design under realistic flow conditions.

The main contributions of this work are: (1) We develop a physically consistent Markov-based state–space model for microfluidic MC channels that integrates diffusion, advection, reversible binding, and flow-out effects within a unified stochastic framework; (2) we derive a discrete-time system model that relates \ac{IM} release and observation processes, yielding closed-form expressions for responses of the microfluidic MC channel; (3) we characterize the channel dynamics under pulse and continuous \ac{IM} release, and derive a closed-form equilibrium gain expression that quantifies the cumulative effects of diffusion, reversible binding, and flow-out processes; and (4) we investigate the impact of flow conditions through the Péclet number and analyze the transition from diffusion-dominated to flow-dominated regimes via numerical simulations.

\section{Preliminary} 
\label{sec:preliminary}

This section establishes the mathematical framework for modeling \ac{IM} propagation in microfluidic MC channels. The spatiotemporal dynamics governed by advection, diffusion, and surface reactions are first described using the advection-diffusion equations. To obtain a tractable system representation, we discretize the channel into a finite set of spatial states and formulate the corresponding discrete-time Markov model. The resulting transition matrix captures the effects of diffusion, advection, binding, and flow-out processes, providing the foundation for the subsequent analysis of equilibrium and transient behaviors. 

\subsection{Mathematical Model}

\begin{figure}
    \centering
    \includegraphics[width=0.6\linewidth]{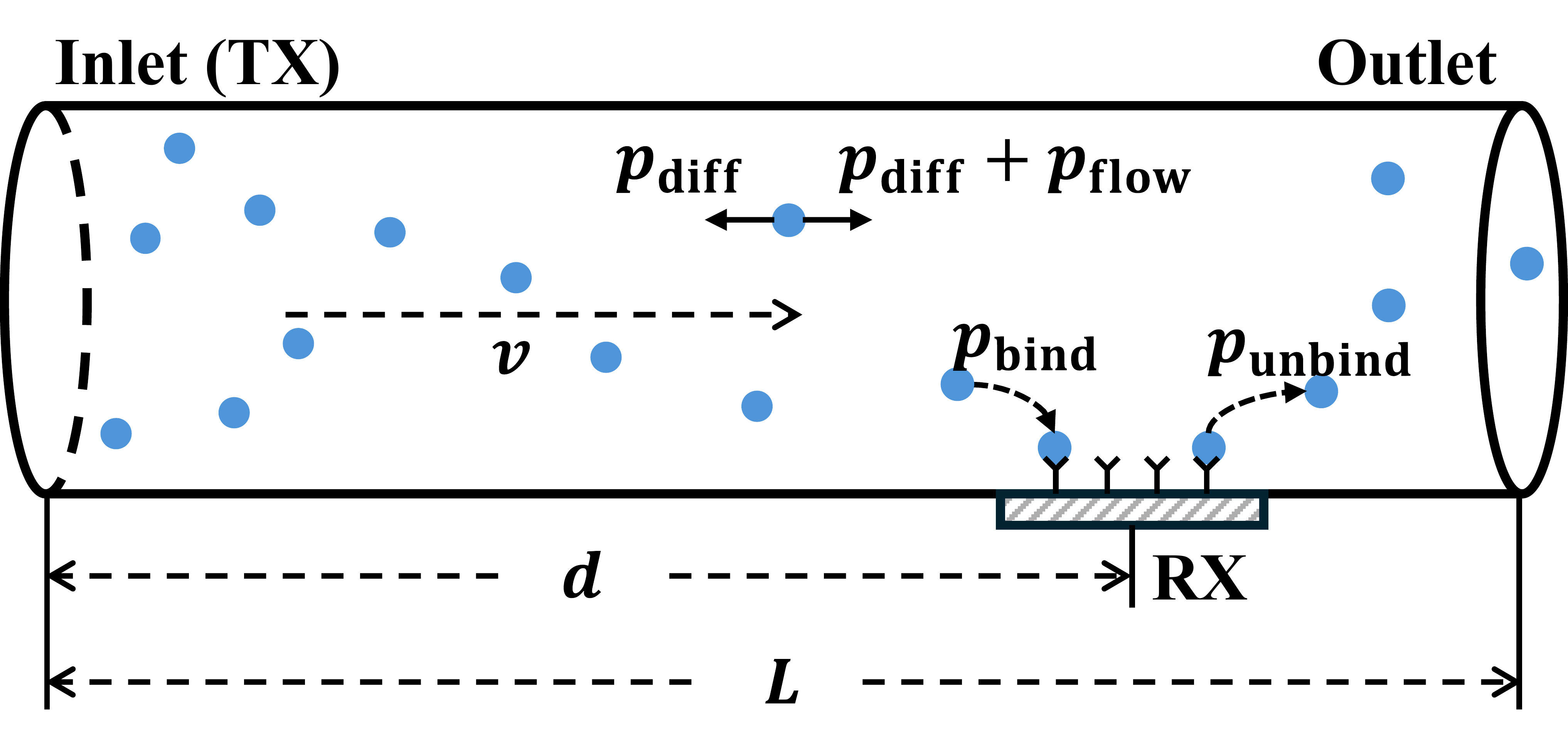}
    \caption{Schematic of the microfluidic channel with a receiving boundary.}
    \label{fig:microfluidic_channel}
    \vspace{-0.5cm}
\end{figure}

We consider a microfluidic MC system in which \acp{IM} are transported within a bounded fluidic domain~$\Omega$. The transmitter releases \acp{IM} into the medium, while the receiver, occupying a reactive surface region~$\Gamma_{\mathrm{RX}}\subset\partial\Omega$, captures them via reversible surface binding, as illustrated in~\cref{fig:microfluidic_channel}. The flow in the microfluidic channel is assumed to be laminar, and IM transport is governed by diffusion, advection, and surface reaction processes at the receiver interface.

Let \( c(\mathbf{r}, t) \) denote the concentration of free \acp{IM} at spatial position \( \mathbf{r}=(x,y,z) \) and time \( t \). The spatiotemporal evolution of \( c \) follows the advection–diffusion equation
\begin{equation}
\frac{\partial c}{\partial t}
= \nabla \!\cdot\! \big( D\,\nabla c - \mathbf{v}\,c \big), 
\quad \mathbf{r}\in\Omega,
\label{eq:cd_eq}
\end{equation}
where \( D \) is the diffusion coefficient, and \( \mathbf{v}=(v_x,v_y,v_z) \) denotes the local flow velocity vector characterizing the microfluidic environment.  

At the reactive boundary~$\Gamma_{\mathrm{RX}}$, IMs undergo reversible binding with immobilized receptor sites (e.g., probe DNAs) following Langmuir-type surface kinetics governed by the association and dissociation rate constants~$k_{\mathrm{on}}$ and~$k_{\mathrm{off}}$, respectively. The temporal evolution of the surface-bound molecular concentration is described by
\begin{equation}
\frac{d c_{\mathrm{b}}}{d t}
= k_{\mathrm{on}}\,c|_{\Gamma_{\mathrm{RX}}}\,(c_{\mathrm{p}}-c_{\mathrm{b}})
- k_{\mathrm{off}}\,c_{\mathrm{b}},
\label{eq:binding_dynamics}
\end{equation}
where $c|_{\Gamma_{\mathrm{RX}}}$ is the local concentration of free IMs at the reactive surface and $c_{\mathrm{p}}$ denotes the total receptor-site concentration. To account for molecular loss due to advection or outflow, a fully absorbing boundary~$\Gamma_{\mathrm{out}}$ is introduced, at which all arriving IMs are immediately removed from the system,
\begin{equation}
c = 0,
\quad \mathbf{r}\in\Gamma_{\mathrm{out}},~t>0,
\label{eq:absorbing_bc}
\end{equation}
representing the \emph{flow-out state} of the microfluidic MC channel. On the remaining non-reactive walls, 
$\Gamma_{\mathrm{ref}} = \partial\Omega \setminus (\Gamma_{\mathrm{RX}} \cup \Gamma_{\mathrm{out}})$, a reflecting (no-flux) boundary condition is imposed,
\begin{equation}
\frac{\partial c}{\partial \mathbf{n}} = 0,
\quad \mathbf{r}\in\Gamma_{\mathrm{ref}},~t>0,
\label{eq:noflux_bc}
\end{equation}
where $\mathbf{n}$ denotes the unit outward normal vector on the boundary surface. This condition indicates that the IMs are completely reflected at the non-reactive boundaries without adsorption or loss.

\subsection{Markov Representation}
\label{subsec:markov_representation}

We model the diffusion–binding dynamics within the microfluidic MC channel as a discrete-time, discrete-space, homogeneous Markov process. This formulation enables efficient computation of ensemble IM behavior via a transition matrix that captures diffusion, advection, and surface reaction effects at each time step.

We divide the channel domain into \( N + 1 \) representative spatial states. The state of a single \ac{IM} at time~$t$ is denoted by \( R(t) \in \mathcal{S} = \{s_1, s_2, \ldots, s_N, s_{N+1}\} \), where \( s_1,\ldots,s_N \) are transient states (free or bound), and \( s_{N+1} \) is the flowed-out state representing irreversible IM loss due to advective flow.

The one-step transition probability from state \( s_j \) to state \( s_i \) within a discrete time step \( \Delta t \) is defined as
\begin{equation}
p(i,j) = \Pr\!\big( R(t+\Delta t) = s_i \,\big|\, R(t) = s_j \big),
\label{eq:transition_def}
\end{equation}
where \( R(t) \) denotes the state of an IM at time \( t \). The overall state evolution is characterized by the transition probability matrix
\begin{equation}
\mathbf{P} = [p(i,j)] \in \mathbb{R}^{(N+1)\times(N+1)}.
\label{eq:P_matrix_def}
\end{equation}
By conservation of probability, each column of $\mathbf{P}$ sums to one, i.e.,
\begin{equation}
\sum_{i=1}^{N+1} p(i,j) = 1, \quad \forall\, j.
\label{eq:prob_conserve}
\end{equation}
The elements of $\mathbf{P}$ encode distinct physical processes in the microfluidic channel: diffusive transport between neighboring spatial states, advective drift toward the outlet, reversible transitions between free and bound states at the receiver surface, and irreversible flow-out from the observation region. In particular, the submatrix $\mathbf{Q}\in\mathbb{R}^{N\times N}$ collects the transition probabilities among transient states, while the vector $\bm{\psi}\in\mathbb{R}^{N\times 1}$ represents the transition probabilities from each transient state to the flow-out state, forming the composite transition matrix as
\begin{equation}
\mathbf{P} =
\begin{bmatrix}
\mathbf{Q} & \mathbf{0}_{N\times1}\\[3pt]
\bm{\psi}^\mathrm{T} & 1
\end{bmatrix}.
\label{eq:P_block_form}
\end{equation}

This discrete Markov representation provides a physically consistent and computationally tractable abstraction of the underlying advection-diffusion dynamics, serving as the foundation for the subsequent system-level analysis of \ac{IM} propagation and observation.

\subsection{Definition of Elementary Transition Probabilities}

To parameterize the transient-state transition matrix~$\mathbf{Q}$, we define four elementary transition probabilities corresponding to the main physical mechanisms in the microfluidic MC channel~\cite{zheng2025DNA-Based}, cf.~\cref{fig:microfluidic_channel}.  

The \emph{diffusion transition probability} is given by
\begin{equation}
    p_{\mathrm{diff}} = \frac{D\,\Delta t}{(\Delta x)^2},
    \label{eq:pdiff}
\end{equation}
representing the probability that an IM diffuses from one spatial state to a neighboring state within one discrete time step~$\Delta t$, where \(D\) is the diffusion coefficient and \(\Delta x\) is the spatial discretization length.

The \emph{binding transition probability} is expressed as
\begin{equation}
    p_{\mathrm{bind}} = k_{\mathrm{on}}\,c_{\mathrm{p}}\,\Delta t,
    \label{eq:pbind}
\end{equation}
and the corresponding \emph{unbinding transition probability} is
\begin{equation}
    p_{\mathrm{unbind}} = k_{\mathrm{off}}\,\Delta t,
    \label{eq:punbind}
\end{equation}
where \(k_{\mathrm{on}}\) and \(k_{\mathrm{off}}\) are the association and dissociation rate constants, respectively, and \(c_{\mathrm{p}}\) is the surface receptor concentration. In this work, we assume \(c_{\mathrm{p}}\!\gg\!c|_{\Gamma_{\mathrm{RX}}}\), so that the receptor concentration \(c_{\mathrm{p}}\) remains nearly constant and both transition probabilities scale linearly with~$\Delta t$.

The \emph{flow-out transition probability} accounting for advective loss is defined as
\begin{equation}
    p_{\mathrm{flow}} = \frac{v\,\Delta t}{\Delta x},
    \label{eq:pflow}
\end{equation}
where \(v\) is the mean flow velocity and quantifies the likelihood that an IM is advected out of the system domain during one time step.

These probabilities determine the non-zero elements of the transient-state transition matrix~$\mathbf{Q}$ and the flow-out vector~$\bm{\psi}$ in the full transition matrix~$\mathbf{P}$; cf.~\cref{eq:P_block_form}. For each transient-state column \(j\), the conservation of probability yields
\begin{equation}
    \sum_{i=1}^{N} Q_{ij} + \psi_j = 1,
    \label{eq:Q_colsum}
\end{equation}
implying
\begin{equation}
    \sum_{i=1}^{N} Q_{ij} = 1 - \psi_j \le 1,
    \label{eq:Q_colsum_le_one}
\end{equation}
where \(Q_{ij}\) is the \((i,j)\)-th element of~\(\mathbf{Q}\), and \(\psi_j\) is the \(j\)-th element of~\(\bm{\psi}\).

\section{System Model} \label{sec:sys_model}
This section develops a Markov-based state–space model for a microfluidic MC channel that captures diffusion, advection, reversible binding, and molecular flow-out processes. 

\subsection{System Overview}

In microfluidic environments, \ac{IM} transport is primarily governed by axial flow, while radial variations are smoothed by diffusion in narrow channels. Hence, the system is modeled as a one-dimensional process along the flow axis. The channel is represented by a one-dimensional Markov chain comprising free, bound, and flow-out states. The model considers a transmitter and a receiver embedded in the channel, forming a single-input single-output (SISO) system. The receptor density on the receiver surface is assumed to be sufficiently high to sustain a stable binding rate. Under laminar flow, the velocity field follows Poiseuille’s law, varying radially from zero at the channel wall to a maximum at the centerline. For analytical tractability, the parabolic velocity profile is approximated by a uniform axial flow—an acceptable assumption for narrow channels or when diffusion effectively smooths radial gradients.\footnote{For wider channels where radial variations become significant, the model can be extended to a two-dimensional Markov process, which is left for future work.}

\subsection{Definition of the Transition Matrix}
\begin{figure}
    \centering
    \includegraphics[width=0.7\linewidth]{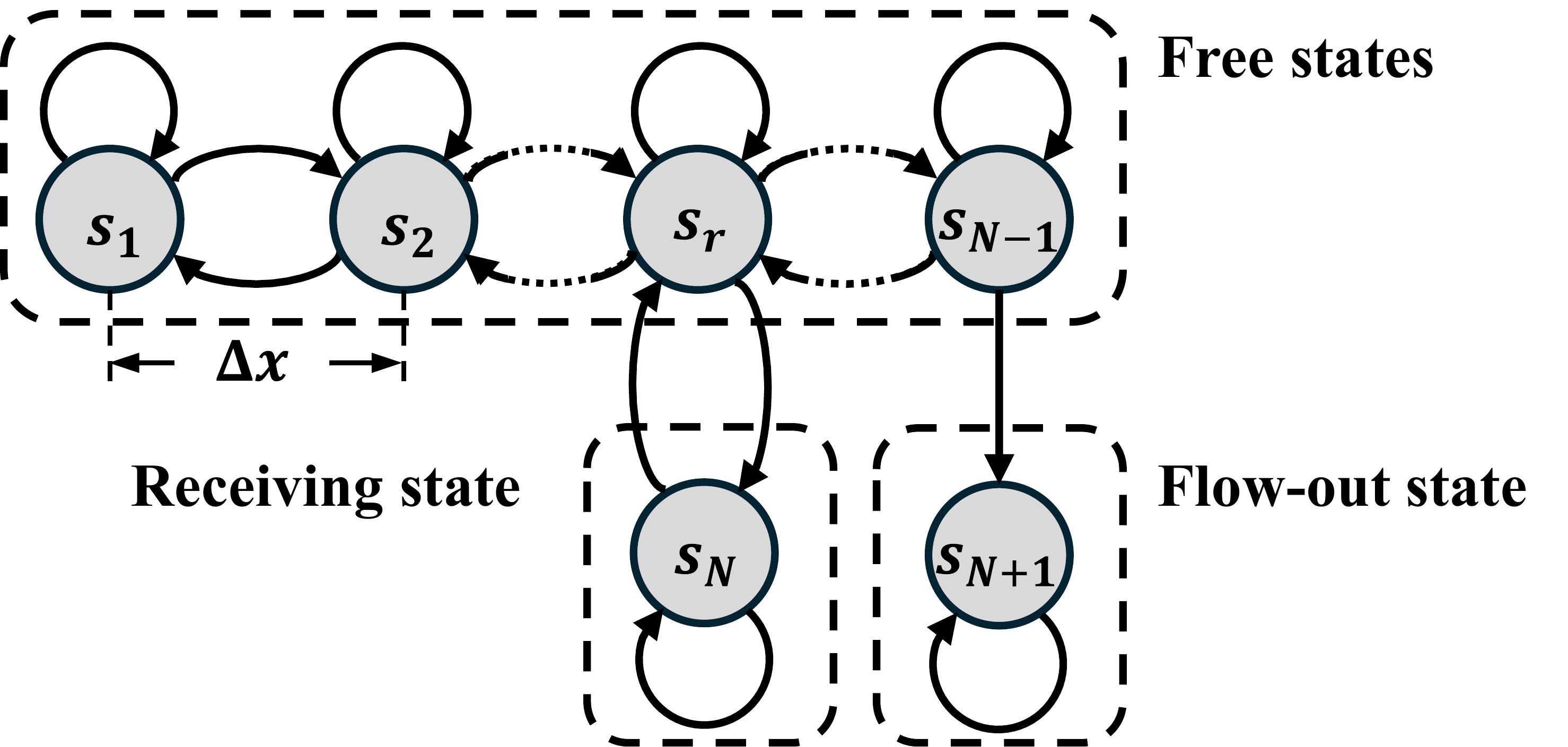}
    \caption{Markov chain of the microfluidic MC channel.}
    \label{fig:markov_chain}
    \vspace{-0.5cm}
\end{figure}
Based on the Markov representation established in~\cref{subsec:markov_representation}, the channel is divided into states \(\{s_1, \ldots, s_N, s_{N+1}\}\), where \(s_1,\ldots,s_{N-1}\) are \emph{free} diffusion states, \(s_N\) is the \emph{bound} state at the receiver surface, and \(s_{N+1}\) is the \emph{flow-out} state representing irreversible molecular loss. The transmitter is located at state~\(s_1\), while the receiver region is represented by the pair of states~\((s_r, s_N)\), where \(s_r\) denotes the free state adjacent to the receiver surface and \(s_N\) represents the corresponding bound state. The effective transmitter–receiver distance is \(d = r\,\Delta x\), and the total channel length is \(L = (N-1)\,\Delta x\), as illustrated in~\cref{fig:markov_chain}.

When the time step~$\Delta t$ is sufficiently small, each IM is assumed to transition only between adjacent states within a single time step, i.e., $p(i,j)=0$ for $|i-j|>1$. According to the definition in~\cref{eq:transition_def}, where $p(i,j)$ denotes the probability of transitioning from state~$s_j$ to state~$s_i$, the non-zero elements of the transition matrix~$\mathbf{P}$ (defined in~\cref{eq:P_block_form}) for the free states $\{s_1, s_2, \ldots, s_{N-1}\}$ can be expressed in terms of the four elementary transition probabilities introduced in~\cref{eq:pdiff,eq:pbind,eq:punbind,eq:pflow} as
\begin{equation}
    p(i{+}1,i) = p_{\mathrm{diff}} + p_{\mathrm{flow}},
    \quad
    p(i{-}1,i) = p_{\mathrm{diff}},
    \label{eq:p_free_states}
\end{equation}
where $p_{\mathrm{diff}}$ and $p_{\mathrm{flow}}$ (defined in~\cref{eq:pdiff,eq:pflow}) represent, respectively, symmetric diffusion between neighboring spatial states and net advective transport toward the downstream direction. Since the microfluidic flow is laminar and predominantly unidirectional, the upstream advective component is neglected.

At the downstream boundary, i.e., between the last free state~$s_{N-1}$ and the flow-out state~$s_{N+1}$, the probability that an IM exits the system during one time step is given by
\begin{equation}
    p(N{+}1, N{-}1) = p_{\mathrm{diff}} + p_{\mathrm{flow}},
    \quad
    p(N{-}1, N{+}1) = 0.
    \label{eq:p_flow_out}
\end{equation}
The flow-out state~$s_{N+1}$ is modeled as an absorbing state with $p(N{+}1,N{+}1)=1$. At the receiver interface, the transition from the free state~$s_r$ to the bound state~$s_N$ and the reverse unbinding transition from~$s_N$ to~$s_r$ occur with probabilities $p_{\mathrm{bind}}$ and $p_{\mathrm{unbind}}$, respectively (defined in~\cref{eq:pbind,eq:punbind}), i.e.,

\begin{equation}
    p(N, r) = p_{\mathrm{bind}},
    \quad
    p(r, N) = p_{\mathrm{unbind}}.
    \label{eq:p_bind_unbind}
\end{equation}

The probability of remaining in the same state, denoted \(p(i,i)\), is computed as the complement of the probabilities of transitions to neighboring states. Specifically, for interior free states \(1 < i < N{-}1\) and \(i \neq r\),
\begin{equation}
    p(i,i) = 1 - p(i{+}1,i) - p(i{-}1,i),
    \label{eq:p_self_interior}
\end{equation}
while for the free state adjacent to the receiver, \(s_r\),
\begin{equation}
    p(r,r) = 1 - p(r{+}1,r) - p(r{-}1,r) - p(N,r).
    \label{eq:p_self_receiver}
\end{equation}
At the reflecting upstream boundary and the bound state, the transition probabilities simplify to
\begin{equation}
    p(1,1) = 1 - p(2,1),
    \quad
    p(N,N) = 1 - p_{\mathrm{unbind}}.
    \label{eq:p_boundaries}
\end{equation}

\subsection{State–Space Channel Model}
\begin{figure}[t]
    \centering
    \includegraphics[width=1.0\linewidth]{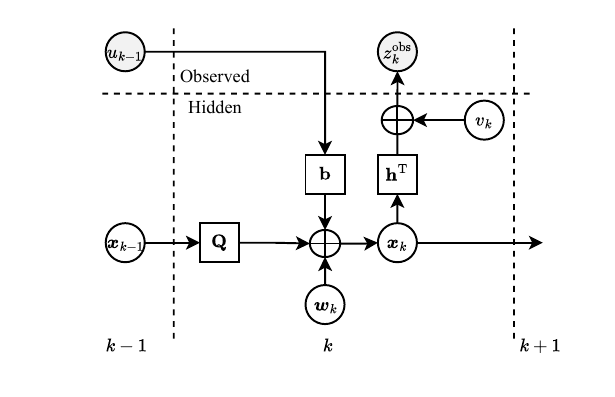}
    \caption{State–space model of the microfluidic MC channel. The hidden Markov state~$\bm{x}_k$ evolves according to the transient-state transition matrix~$\mathbf{Q}$, driven by the input~$\mathbf{b}u_{k-1}$ and the process noise~$\mathbf{w}_k$. The observation~$z_k^{\mathrm{obs}}$ is obtained from the state~$\bm{x}_k$ through the observation vector~$\mathbf{h}$, with the observation noise~$v_k$ accounting for measurement uncertainty.} 
    \label{fig:systematic_graph}
    \vspace{-0.5cm}
\end{figure}
The vector $\bm{x}_k = [x_{1,k}, \dots, x_{N,k}]^\mathrm{T}$ represents the numbers of~\acp{IM} in the transient states at discrete time step~$k$. Each element $x_{i,k}$ denotes the expected number of~\acp{IM} in the $i$-th transient state (either free or bound). The dynamics of the single-input single-output (SISO) system are governed by:
\begin{equation}
    \bm{x}_k = \mathbf{Q}\bm{x}_{k-1} + \mathbf{b}\,u_{k-1},
    \label{eq:state_evolution}
\end{equation}
where $u_{k-1}$ denotes the \emph{input}, i.e., the number of IMs released by the transmitter at time step~$k{-}1$, and $\mathbf{b}\in\mathbb{R}^{N\times1}$ is the input coupling vector that determines how the newly released IMs are initially distributed among the transient states. In this work, $\mathbf{b} = [1, 0, \ldots, 0]^\mathrm{T}$ since the transmitter is located at state~$s_1$.

The observable \emph{output} of the system corresponds to the number of bound IMs detected at the receiver surface. The observation at time step~$k$ is expressed as
\begin{equation}
    z_k^{\mathrm{obs}} = \mathbf{h}^\mathrm{T}\bm{x}_k,
    \label{eq:obs_eq}
\end{equation}
where $\mathbf{h}\in\mathbb{R}^{N\times 1}$ is the \emph{observation vector} that maps the numbers of \acp{IM} in the transient states to the measurable bound states at the receiver.\footnote{Noise effects are neglected in this work and left for future investigation.}

The cumulative number of IMs that have flowed out of the system evolves according to
\begin{equation}
    z_k^{\mathrm{out}} = z_{k-1}^{\mathrm{out}} + \bm{\psi}^\mathrm{T}\bm{x}_{k-1},
    \label{eq:out_eq}
\end{equation}
where $\bm{\psi}\in\mathbb{R}^{N\times 1}$ represents the transition coefficients from the transient states to the flow-out state, and the term $\bm{\psi}^\mathrm{T}\bm{x}_{k-1}$ quantifies the \ac{IM} flux leaving the system during time step~$k$. \Cref{eq:state_evolution,eq:obs_eq,eq:out_eq} together describe the state–space model of the proposed microfluidic MC channel, as illustrated in~\Cref{fig:systematic_graph}. By iteratively applying~\cref{eq:state_evolution}, the transient-state vector at time step~$k$ is obtained as
\begin{equation}
    \bm{x}_k
    = \mathbf{Q}^k\bm{x}_0
    + \sum_{i=0}^{k-1}\mathbf{Q}^{\,i}\mathbf{b}\,u_{k-i-1},
    \label{eq:iterative_form}
\end{equation}
where $\bm{x}_0$ denotes the initial numbers of IMs in the transient states. The first term $\mathbf{Q}^k\bm{x}_0$ captures the natural decay of the initial IMs under the transition dynamics, while the summation term $\sum_{i=0}^{k-1}\mathbf{Q}^{\,i}\mathbf{b}\,u_{k-i-1}$ accumulates the contributions of past inputs weighted by the channel memory~$\mathbf{Q}^i$.

Substituting~\cref{eq:iterative_form} into~\cref{eq:obs_eq} yields a convolutional form of the observation:
\begin{equation}
    z_k^{\mathrm{obs}}
    = \mathbf{h}^\mathrm{T}\mathbf{Q}^k\bm{x}_0
    + \sum_{i=0}^{k-1} g_i\,u_{k-i-1},
    \label{eq:obs_conv}
\end{equation}
where
\begin{equation}
    g_i \triangleq \mathbf{h}^\mathrm{T}\mathbf{Q}^i\mathbf{b}
    \label{eq:cir_def}
\end{equation}
denotes the \emph{channel impulse response}~(CIR) of the microfluidic MC channel. The term~$g_i$ quantifies the expected number of IMs detected at the receiver at time step~$i$ in response to a unit molecular pulse emitted by the transmitter at time~$0$. 

\section{Channel Dynamics and Characteristics} \label{sec:channel_dynamics}

The characteristics of the proposed microfluidic MC are determined by the transient-state transition matrix~$\mathbf{Q}$, which governs the temporal evolution of the IM population within the transient states. To illustrate different aspects of the channel dynamics, we consider two representative input patterns: \emph{pulse release}, which highlights the transient response, and \emph{continuous release}, which characterizes the equilibrium behavior. 

For simplicity, we assume that no IMs are initially present in the channel, i.e., $\bm{x}_0 = \mathbf{0}$. We further focus on a SISO configuration, where the transmitter releases a scalar input~$u_k$ and the receiver measures a scalar output~$z_k^{\mathrm{obs}}$. In this case, the observation model~\cref{eq:obs_conv} reduces to
\begin{equation}
    z_k^{\mathrm{obs}} = \sum_{i=0}^{k-1} g_i\,u_{k-i-1},
    \label{eq:siso_conv}
\end{equation}
where $g_i = \mathbf{h}^\mathrm{T}\mathbf{Q}^{\,i}\mathbf{b}$ is the scalar CIR at time step~$i$.

\subsection{Pulse Release}

We first consider a pulse-release scenario in which the transmitter emits an instantaneous burst of IMs at time~$k=0$. The input signal is modeled as a discrete delta function,
\begin{equation}
    u_k = u_0\,\delta[k],
    \label{eq:pulse_input}
\end{equation}
where $u_0$ denotes the number of IMs released during the pulse. Substituting~\cref{eq:pulse_input} into~\cref{eq:siso_conv} yields
\begin{equation}
    z_k^{\mathrm{obs}} = g_{k-1}\,u_0,
    \label{eq:pulse_output}
\end{equation}
where $g_{k-1}$ represents the CIR that captures the transient diffusion–binding dynamics following an instantaneous IM release.

\subsection{Continuous Release}

Next, we consider a continuous release scenario in which the transmitter emits IMs at a constant rate over time. In this case, the input sequence is modeled as
\begin{equation}
    u_k = u_0, \quad k \ge 0,
    \label{eq:continuous_input}
\end{equation}
where $u_0$ denotes the constant number of IMs released per time step. Substituting~\cref{eq:continuous_input} into~\cref{eq:siso_conv} gives
\begin{equation}
    z_k^{\mathrm{obs}} = \left(\sum_{i=0}^{k-1} g_i\right) u_0,
    \label{eq:continuous_output}
\end{equation}
where the partial sum $\sum_{i=0}^{k-1} g_i$ represents the discrete-time step response of the MC channel up to time~$k$.

As $k \to \infty$, the system approaches its equilibrium, and the total observed IM concentration converges to  
\begin{equation}
    z_{\infty}^{\mathrm{obs}} = \left(\sum_{i=0}^{\infty} g_i\right) u_0.
    \label{eq:equilibrium_output}
\end{equation}
According to the Perron-Frobenius theorem~\cite{levin2017markov}, the spectral radius of the overall transition matrix satisfies $\rho(\mathbf{P}) \le 1$. 
Since the submatrix $\mathbf{Q}$ corresponds to the transient states (cf. ~\cref{eq:Q_colsum_le_one}), it follows that $\rho(\mathbf{Q}) < 1$. Hence, the Neumann series $\sum_{i=0}^{\infty} \mathbf{Q}^{\,i}$ converges to $(\mathbf{I} - \mathbf{Q})^{-1}$, yielding  
\begin{equation}
    \sum_{i=0}^{\infty} g_i
    = \mathbf{h}^\mathrm{T}\!\left(\sum_{i=0}^{\infty}\mathbf{Q}^{\,i}\right)\!\mathbf{b}
    = \mathbf{h}^\mathrm{T}(\mathbf{I}-\mathbf{Q})^{-1}\mathbf{b}.
    \label{eq:equilibrium_sum}
\end{equation}

Hence, the equilibrium output can be expressed as
\begin{equation}
    z_{\infty}^{\mathrm{obs}}
    = \mathbf{h}^\mathrm{T}(\mathbf{I}-\mathbf{Q})^{-1}\mathbf{b}\,u_0,
    \label{eq:equilibrium_gain}
\end{equation}
which represents the equilibrium gain of the microfluidic MC channel. Physically, the term $\mathbf{h}^\mathrm{T}(\mathbf{I}-\mathbf{Q})^{-1}\mathbf{b}$ quantifies the cumulative influence of diffusion, reversible binding, and flow-out processes over infinite time, describing how released IMs are gradually distributed and detected at the receiver surface.

\section{Numerical Results}\label{sec:numerical_results}
\begin{figure*}[t]
    \centering
    \subfigure[Pulse release.]{
        \label{fig:pulse_release}
        \centering
        \includegraphics[width=0.45\linewidth]{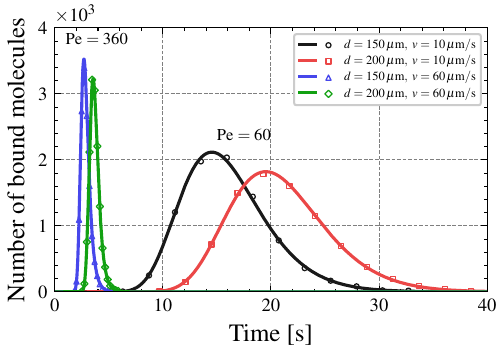}
    }
    \hfill
    \subfigure[Continuous release.]{
        \label{fig:continuous_release}
        \centering
        \includegraphics[width=0.45\linewidth]{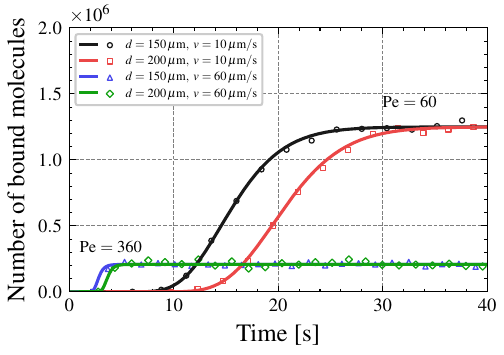}
    }
    \vspace{-0.3cm}
    \caption{Observed signal~$z_k^{\mathrm{obs}}$ under (a) pulse release and (b) continuous release of IMs for different transmitter–receiver distances~$d$ and flow velocities~$v$. Solid lines show the analytical model from~\cref{eq:pulse_output,eq:continuous_output}, while markers denote particle-based simulation (PBS) results. Simulation parameters: (a)~$10^5$~IMs are instantaneously emitted at $t=0$ (\(u_0 = 10^5\)) according to~\cref{eq:pulse_input}; (b)~$5\times10^3$~IMs are continuously injected per~$\Delta t$ (\(u_0 = 10^3\)) following~\cref{eq:continuous_input}.}
    \label{fig:observation}
    \vspace{-0.6cm}
\end{figure*}
\begin{table}[t]
    \centering
    \caption{Simulation Parameters.}
    \label{tab:parameters}
    \renewcommand{\arraystretch}{1.2}
    \begin{tabular}{@{}lccc@{}}
        \toprule
            \textbf{Parameter}          & \textbf{Symbol}   & \textbf{Value}    & \textbf{Unit} \\ 
            \midrule
            Diffusion coefficient       & $D$               & \num{5e-11}       & \si{\metre\squared\per\second} \\
            Concentration of receptors  & $c_\text{p}$      & \num{1e-8}        & \si{M} \\
            Number of states            & $N$               & \num{301}         & -- \\
            Binding rate                & $k_\text{on}$     & \num{6e8}         & \si{ M^{-1}.s^{-1}} \\
            Unbinding rate              & $k_\text{off}$    & \num{3}           & \si{s^{-1}} \\
            Spatial step size           & $\Delta x$        & \num{1e-8}        & \si{\meter} \\
            Time step size              & $\Delta t$        & \num{8e-4}        & \si{s} \\
        \bottomrule
    \end{tabular}
    \vspace{-0.5cm}
\end{table}
In this section, we examine the response of the proposed microfluidic MC communication system under different flow conditions and release schemes to characterize the channel dynamics. Both pulse and continuous IM releases are considered across a range of Péclet numbers ($\mathrm{Pe}=vL/D$), which quantify the relative strength of advection and diffusion. All simulations are performed in MATLAB using the parameters summarized in~\cref{tab:parameters}, unless stated otherwise. 

In the pulse release scenario, a finite number of IMs is instantaneously emitted at $t=0$, and the temporal evolution of the received concentration characterizes the CIR. In the continuous release scenario, IMs are released at a constant rate, and the equilibrium concentration at the receiver is analyzed for different $\mathrm{Pe}$ values. \Cref{fig:observation} illustrates the corresponding observed signals~$z_k^{\mathrm{obs}}$ for both release types, as discussed in~\cref{sec:channel_dynamics}. PBS results agree closely with the proposed system model—\cref{eq:pulse_output} for pulse release and~\cref{eq:continuous_output} for continuous release—validating the model’s accuracy in capturing both transient and equilibrium behaviors.

\Cref{fig:pulse_release} illustrates the observed signal~$z_k^{\mathrm{obs}}$ under pulse release of \acp{IM} for different Péclet number scenarios. For moderate flow conditions ($\mathrm{Pe}=60$, $v=10~\mu\text{m/s}$), the bound-IM response exhibits a broad temporal distribution with a delayed peak due to diffusion-dominant transport. In contrast, for higher flow rates ($\mathrm{Pe}=360$, $v=60~\mu\text{m/s}$), IMs are advected more rapidly toward the receiver, producing an earlier and sharper transient peak with a higher amplitude. Across both $\mathrm{Pe}$ conditions, increasing the transmitter–receiver distance leads to a longer propagation delay and a lower peak amplitude. After the transient response, the bound-IM concentration decays to zero, indicating the complete clearance of IMs from the system.

\Cref{fig:continuous_release} depicts the observed signal~$z_k^{\mathrm{obs}}$ under continuous release of \acp{IM} for different Péclet number scenarios. For the diffusion-dominant case ($\mathrm{Pe}=60$, $v=10~\mu\text{m/s}$), the number of bound IMs increases monotonically and eventually stabilizes at a equilibrium level (cf.~\cref{eq:equilibrium_gain}), where the binding and unbinding fluxes reach dynamic equilibrium. In contrast, for the advection-dominant case ($\mathrm{Pe}=360$, $v=60~\mu\text{m/s}$), the rapid flow shortens the residence time of IMs in the channel, limiting their binding probability and leading to a significantly lower equilibrium concentration. Moreover, increasing the transmitter–receiver distance reduces the effective IM flux and delays the approach to equilibrium.

Overall, the results demonstrate that the proposed model consistently captures the transient response to impulsive excitation and the equilibrium behavior under sustained input. Higher flow velocity (larger $\mathrm{Pe}$) accelerates the transient dynamics in the pulse-release case but suppresses the equilibrium binding level in the continuous-release case, revealing a fundamental trade-off between response speed and binding efficiency in microfluidic MC channels.

\section{Conclusion}
 This paper presents a Markov-based system model for microfluidic MC channels. By discretizing the advection-diffusion dynamics, the proposed model provides a compact and physically consistent representation of diffusion, flow, and reversible binding processes. Numerical results demonstrate that the model accurately captures both transient and equilibrium behaviors under various flow conditions. The proposed framework establishes a tractable foundation for future studies on channel estimation, detection, and system optimization in microfluidic MC. Future work will incorporate noise modeling and extend the framework toward a complete MC system design. Moreover, multiple-input multiple-output (MIMO) configurations for nanonetworks will be explored to investigate spatial diversity and inter-channel coupling effects.
 \vspace{-0.4cm}

\bibliographystyle{IEEEtran}
\bibliography{references}

\end{document}

%% file: preamble.tex
\usepackage{amsmath, amssymb, amsfonts, bm, mathtools}
\usepackage{amsthm}

\usepackage[version=4]{mhchem}
\usepackage{chemfig}

\usepackage{algorithm}
\usepackage{algorithmic}

\usepackage{graphicx}
\usepackage{subfigure}
\usepackage{svg}
\usepackage{tikz}
\usepackage[american]{circuitikz}
\usetikzlibrary{
  arrows.meta, 
  decorations.markings, 
  circuits, 
  automata,
  arrows, 
  positioning, 
  calc
}

\usepackage{booktabs}
\usepackage{makecell}

\usepackage{textcomp}
\usepackage{soul}       
\usepackage{verbatim}
\usepackage{xcolor} 

\usepackage{cite}
\usepackage[hidelinks]{hyperref}
\usepackage[capitalise,noabbrev]{cleveref}
\Crefformat{figure}{#2Fig.~#1#3}
\Crefformat{equation}{#2Eq.~(#1)#3}
\crefname{equation}{Eq.}{Eqs.}
\crefname{figure}{Fig.}{Figs.}

\usepackage{xr}
\usepackage{xr-hyper} 

\usepackage[
  range-phrase=--,
  per-mode=symbol-or-fraction,
  range-units=single,
  list-units=single,
  detect-all,
  list-final-separator={, and },
]{siunitx}

\usepackage{acro}
\usepackage{mfirstuc}

\DeclareAcronym{MC}{short=MC, long=molecular communication}
\DeclareAcronym{MCvD}{short=MCvD, long=molecular communication via diffusion}
\DeclareAcronym{IM}{short=IM, long=information molecule, short-plural=s, long-plural=s}
\DeclareAcronym{TX}{short=TX, long=transmitter, short-plural=s, long-plural=s}
\DeclareAcronym{RX}{short=RX, long=receiver, short-plural=s, long-plural=s}
\DeclareAcronym{IoBNT}{short=IoBNT, long=Internet of Bio-Nano Things}
\DeclareAcronym{ssDNA}{short=ssDNA, long=single-stranded DNA}
\DeclareAcronym{cDNA}{short=cDNA, long=complementary DNA}

\DeclareAcronym{EM}{short=EM, long=electromagnetic}
\DeclareAcronym{FSO}{short=FSO, long=free-space optical}

\DeclareAcronym{ML}{short=ML, long=maximum-likelihood}
\DeclareAcronym{LS}{short=LS, long=least-squares}
\DeclareAcronym{NNLS}{short=NNLS, long=non-negative least squares}
\DeclareAcronym{MAP}{short=MAP, long=maximum a posteriori}
\DeclareAcronym{MMSE}{short=MMSE, long=minimum mean squared error}
\DeclareAcronym{MIMO}{short=MIMO, long=multiple-input multiple-output}
\DeclareAcronym{CDMA}{short=CDMA, long=code division multiple access}

\DeclareAcronym{MSE}{short=MSE, long=mean squared error}
\DeclareAcronym{FIM}{short=FIM, long=Fisher information matrix}
\DeclareAcronym{CRB}{short=CRB, long=Cramér-Rao bound}

\DeclareAcronym{SNR}{short=SNR, long=signal-to-noise ratio}
\DeclareAcronym{SINR}{short=SINR, long=signal-to-interference-plus-noise ratio}

\usepackage{standalone}

\usepackage{listofitems} 
\usepackage[outline]{contour} 
\contourlength{1.4pt}

\colorlet{myred}{red!80!black}
\colorlet{myblue}{blue!80!black}
\colorlet{mygreen}{green!60!black}
\colorlet{myorange}{orange!70!red!60!black}
\colorlet{mydarkred}{red!30!black}
\colorlet{mydarkblue}{blue!40!black}
\colorlet{mydarkgreen}{green!30!black}

\usepackage{titlesec}
\titlespacing\section{0pt}{6pt plus 2pt minus 2pt}{4pt plus 2pt minus 2pt}
\titlespacing\subsection{0pt}{4pt plus 2pt minus 2pt}{3pt plus 2pt minus 2pt}